\documentclass{iau} 
\usepackage{graphicx}
\usepackage{natbib}
\usepackage{hyperref}
\usepackage{subfig}
\title[BA-type supergiants in NGC300 with MUSE] %% give here short title %%
{MUSE 3D spectroscopy of BA-type supergiants in NGC 300
}
\author[Gemma Gonz\'alez-Tor\`a \textit{et al.}]   %% give here short author list %%
{Gemma Gonz\'alez-Tor\`a$^{1,2,3}$,
 Miguel A.~Urbaneja$^{3}$,
 Norbert Przybilla$^{3}$,
 Stefan Dreizler$^{4}$,
 Martin M.~Roth$^{5}$,
 Sebastian Kamann$^{2}$,
 \and Norberto Castro$^{5}$}

\affiliation{$^1$European Southern Observatory (ESO), Karl-Schwarzschild-Str. 2, 85748 Garching bei München, Germany
\\[\affilskip]
$^{2}$Astrophysics Research Institute, LJMU, 146 Brownlow Hill, Liverpool L3 5RF, UK
\\[\affilskip]
$^3$Institut für Astro- und Teilchenphysik, Universität Innsbruck, Technikerstr. 25/8, 6020 Innsbruck, Austria
\\[\affilskip]
$^4$Institute for Astrophysics, University of Göttingen, Friedrich-Hund-Platz 1, 37077 Göttingen, Germany
\\[\affilskip]
$^5$Leibniz-Institut für Astrophysik (AIP), An der Sternwarte 16, 14482 Postdam, Germany
}

\pubyear{2022}
\volume{361}  %% insert here IAU Symposium No.
\jname{Massive Stars Near and Far}
\editors{N.~St-Louis, J.~S.~Vink \& J.~Mackey, eds.}
\begin{document}

\maketitle

\begin{abstract}
We present the results obtained using spectroscopic data taken with the intermediate-resolution Multi Unit Spectroscopic Explorer (MUSE) of B and A-type supergiants and bright giants in the Sculptor Group galaxy NGC 300. For our analysis, a hybrid local thermodynamic equilibrium (LTE) line-blanketing+non-LTE method was used to improve the previously published results for the same data. In addition, we present some further applications of this work, which includes extending the flux-weighted gravity luminosity relationship (FGLR), a distance determination method for supergiants. This pioneering work opens up a new window to explore this relation, and also demonstrates the enormous potential of integral field spectroscopy (IFS) for extragalactic quantitative stellar studies.
%We present an overview of PION, an open-source software project for solving radiation-magnetohydrodynamics equations on a nested grid, aimed at modelling asymmetric nebulae around massive stars.
%A new implementation of hybrid OpenMP/MPI parallel algorithms is briefly introduced, and improved scaling is demonstrated compared with the current release version.
%Three-dimensional simulations of an expanding nebula around a Wolf-Rayet star are then presented and analysed, similar to previous 2D simulations in the literature.
%The evolution of the emission measure of the gas and the X-ray surface brightness are calculated as a function of time, and some qualitative comparison with observations is made.

\begin{keywords}
Galaxies: individual (NGC 300) – Galaxies: distances and redshifts – Stars: early-type – Stars: fundamental parameters
– supergiants
\end{keywords}
\end{abstract}

\firstsection % if your document starts with a section,
\section{Introduction}
%%%%%%%%%%%%%%%%%%%%%%%%%%%%%%%%%%%%%%%%%%%%%%%%%%%%%%%%%%
Resolving individual stars in other galaxies has been a turning point for astronomy \citep[e.g.,][]{1929ApJ....69..103H,Baade44}. The best candidates for such observations are massive BA-type supergiants, as they are the brightest objects in optical light, reaching
absolute visual magnitudes of $M_V$\,$\simeq$\,$-9.5$
\citep[e.g.][]{1979ApJ...232..409H,1987AJ.....94.1156H}. The spiral galaxy NGC\,300 is located at the Sculptor Group, close to the Galactic southern pole. These galaxies are the least affected by Galactic extinction and therefore convenient targets to study. Moreover, NGC\,300 is oriented face-on, being a great candidate for quantitative multi-object slit spectroscopy \citep[e.g.,][]{2002ApJ...567..277B,2004ApJ...600..182B,2005ApJ...622..862U}.
An extremely powerful tool for this purpose is the so-called integral field spectroscopy (IFS). This technique allows for a spectrum for each pixel across an image to be obtained simultaneously. The Multi Unit Spectroscopic Explorer \citep[MUSE,][]{2014Msngr.157...13B} on
the European Southern Observatory Very Large Telescope (ESO VLT) is groundbreaking in this context, as it combines a wide field of view with high spatial
sampling. Spectroscopic analysis of the brightest objects in nearby galaxies, i.e. supergiant stars, can be performed to determine their stellar parameters. 
%Some of the previous work on NGC300 supergiants may be cited/briefly mentioned:
%Bresolin et al. 2002, on the earlier FORS observations of BSGs
%Bresolin et al. 2004, on the negligible effect of photometric variations of the stars on the FGLR
%Urbaneja et al. 2005, on alpha-element abundance gradients, of what can be done at higher S/N

%maybe also  Urbaneja et al. 2017 on the LMC FGLR

\begin{figure}
\centering
\includegraphics[width=.7\linewidth]{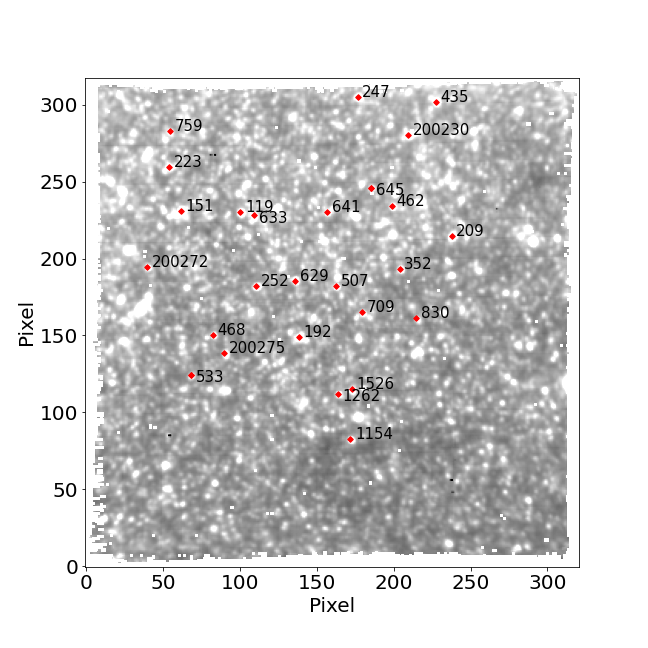}
\caption{Chart of NGC\,300 field (i) in spaxel coordinates with the programme stars marked in red and their ID numbers marked in black (according to Paper~I).
%The image is stacked over all recorded wavelengths in the datacube.
}
\label{fig:photometry}
\end{figure}

The present study based on \citet{2022A&A...658A.117G} provides a detailed analysis of MUSE
spectra of BA-type supergiants and bright giants in NGC\,300 in the field seen in Figure~\ref{fig:photometry} from \citep[][henceforth Paper~I]{2018A&A...618A...3R}. The analysis is based on synthetic spectra accounting for deviations from local thermodynamic equilibrium (LTE).

%Introductory comments, citing literature on expanding wind bubbles that are often radiative and subject to dynamical instabilities \citep{GarMac95b}, and effects of thermal conduction \citep{ComKap98, MeyMacLan14} and magnetic fields \citep{VanMelMar15, MeyMigKui17} can be significant.

%%%%%%%%%%%%%%%%%%%%%%%%%%%%%%%%%%%%%%%%%%%%%%%%%%%%%%%%%%
\section{Methods}
%%%%%%%%%%%%%%%%%%%%%%%%%%%%%%%%%%%%%%%%%%%%%%%%%%%%%%%%%%
%observations 
The spectroscopic data were obtained using MUSE
\citep{2014Msngr.157...13B}, in  
the wide field mode (WFM) with 1$'\times$1$'$~spatial coverage 
and 0.2$''$ sampling. The pointing observed under the best seeing conditions
(FWHM\,=\,0.47$''$-0.59$''$, measured from the data) was investigated.

The initial reduction was achieved with the MUSE pipeline V1.0
\citep[see Paper~I for details][]{2020A&A...641A..28W}. 
The final data were produced in the form of a datacube, 
and the spectra of 606 individual sources were extracted using 
the PampelMUSE software \citep{kamann13}.

Out of these 606 extracted sources, 26 were classified as
late-B to early-A supergiants or bright giants in Paper~I with a 7\,$<$\,S/N\,$\lesssim$\,20, the minimum was demanded for our analysis to provide valid results. They are identified in 
Fig.~\ref{fig:photometry}. Due to model restrictions, we ended up with 16 objects that fulfilled the criteria for a quantitative analysis \citep[see][for more details]{2022A&A...658A.117G}.

%For two stars, no Johnson photometric data are available. Four objects show, in their spectra, emission lines that are characteristic of low-excitation H\,{\sc ii} regions, such as [N\,{\sc ii}] and [S\,{\sc ii}], and very weak [O\,{\sc iii}]. The stellar hydrogen lines are therefore expected to be contaminated by nebular emission. Moreover, the stars are too cool to excite the nebulae; therefore, it is likely that nearby OB-type stars contribute to the recorded spectra. Finally, four more stars will later turn out to have atmospheric parameters that lie beyond those covered by our model grids. Consequently, only 16 objects fulfilled the criteria for a quantitative analysis, which is discussed in the following.
 
%model and analysis
We consider a grid of model atmospheres covering the parameter space defined by the effective temperature  $T_{\mathrm{eff}}$ and surface gravity $\log g$. We adopt the modelling methodology by \citet{Norbert06} for the analysis of our final sample 
of 16 stars. Very briefly, the method employs a combination of model atmosphere structures calculated under the assumption 
of LTE+line-blanketing and a detailed non-LTE level population as well as line-formation 
calculations. The reader is referred to \citet{Norbert06} for the advantages and drawbacks of this hybrid approach, as well as its limitations. 
%\citep{Kudritzki08,2015ApJ...805..182G}, 
%constrains the metallicity at an average $[Z]$\,=\,$-$0.2 dex. Furthermore, following the results obtained in the high resolution work by \cite{Przybilla02}, we adopted a dependency of the microturbulence with $T_{\mathrm{eff}}$ and $\log g$ 

\begin{figure}
\centering
\includegraphics[width=.8\linewidth]{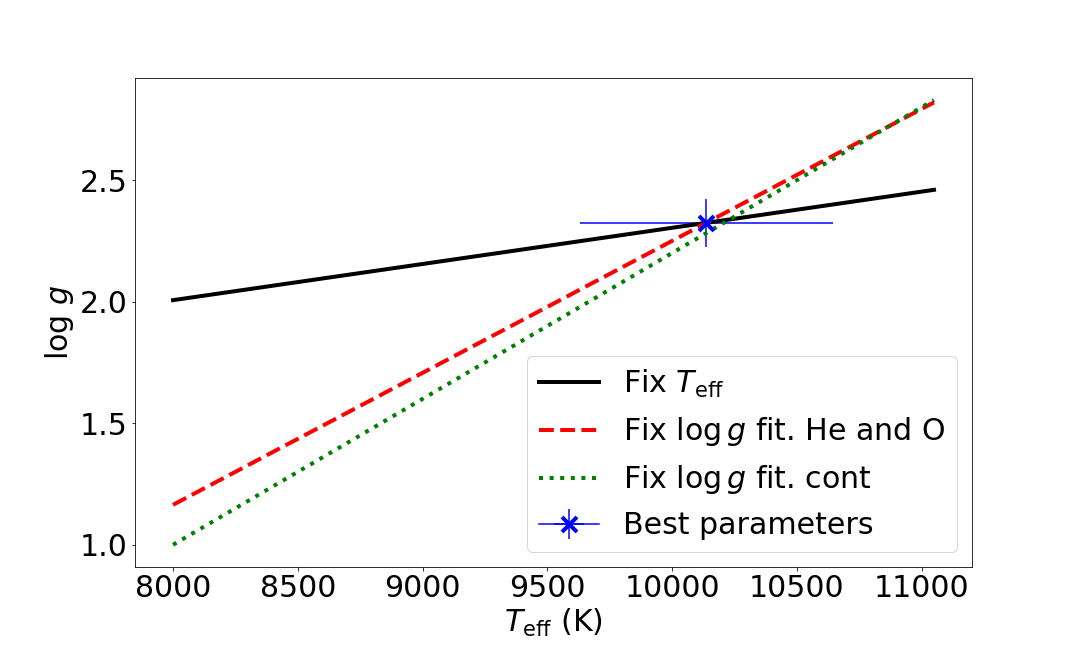}
\caption{Example of the method used to determine the temperature and surface gravity for star \#151. The solid black line shows the locus obtained by varying $\log g$ (in cgs units) for each fixed model-grid $T_\mathrm{eff}$ fitting the hydrogen lines. As red-dashed and green-dotted lines, we show the loci obtained by varying $T_\mathrm{eff}$ at each fixed model-grid $\log g$ best fitting the He\,{\sc i} and O\,{\sc i} lines, and the metal-line dominated  4950-5600\,{\AA} region, respectively. The blue cross marks the intersection of the loci, corresponding with the adopted atmospheric parameter values.
\label{fig:id151}}
\end{figure}

We followed a well-established methodology to find the best solution for each object. %in the $T_{\mathrm{eff}}$--$\log\,\mathrm{g}$ plane. 
 First, by adopting $T_{\mathrm{eff}}$, we found the model that best reproduces the features sensitive to gravity changes
 (the hydrogen Balmer lines); this step was repeated for different adopted $T_{\mathrm{eff}}$ values, hence allowing us to define the locus of models for which the Balmer lines are equally well represented.  This is referred as the $\log g$ locus. Similarly, but adopting the surface gravity, we identified the model that best reproduces the $T_{\mathrm{eff}}$ sensitive features (in our case either the metal lines He\,{\sc i} and O\,{\sc i} lines, or the metal-line dominated  4950-5600\,{\AA} region). We repeated this step for different values of $\log g$, defining the $T_{\mathrm{eff}}$ locus. Finally, the intersection of both lines represents the best possible solution for a given object in the  $T_{\mathrm{eff}}$--$\log g$ plane (see Fig.\,\ref{fig:id151}).
 
A catalogue of Johnson $B$- and $V$-band magnitudes for sources in NGC\,300, based on the work by \citet{Piet01}, was kindly provided by F.~Bresolin (private communication). It allowed to calculate bolometric magnitudes $M_\mathrm{bol}$ from the extinction-corrected apparent $V$-band magnitudes, the distance to NGC\,300 of $d$\,=\,1.86$\pm$0.07 Mpc as determined by \cite{Rizzi2006}., and the bolometric corrections ($B.C.$) for each object individually, calculated from tailored models.

%The source code and tutorials for PION v2.0 are available from a website\footnote{\href{https://www.pion.ie/}{https://www.pion.ie}} with a BSD 3-clause license.
%The most difficult work involved making various function calls threadsafe for OpenMP execution, which required some re-design of the data structures.
%Code optimization work is still ongoing, but there is already signficiant improvement.
%Fig.~\ref{fig:scaling} shows preliminary results of the strong scaling using upgraded MPI communication and either 1 or 5 OpenMP threads per MPI process.

%%%%%%%%%%%%%%%%%%%%%%%%%%%%%%%%%%%%%%%%%%%%%%%%%%%%%%%%%%
\section{The flux-weighted gravity luminosity relationship (FGLR)}\label{sec:fglr}
%%%%%%%%%%%%%%%%%%%%%%%%%%%%%%%%%%%%%%%%%%%%%%%%%%%%%%%%%%
The FGLR was first derived by \citet{2003ApJ...582L..83K} as a new method for distance determination of supergiants:
\begin{equation}
\label{eq:fglrfinal}
-M_{\mathrm{bol}}=a(\log g_{\mathrm{F}}-1.5)+b
\end{equation}
with $a$ determined by the mass-luminosity relation ($L$\,$\propto$\,$M^x$) exponent $x$ to $a$\,=\,2.5$x/(1-x)$, and $g_{\mathrm{F}}$ the so-called flux-weighted gravity, defined as $g_{\mathrm{F}}$\,=\,$g\times\left(T_{\mathrm{eff}}/10^{4}\right)^{-4}$. %For massive stars, $x$\,$\approx$\,3 is found.
%\begin{equation}
%\label{eq:a}
%a=2.5x/(1-x).
%\end{equation}

This relation holds for all supergiants and bright giants that have a constant luminosity track when they move to the right of the Hertzsprung-Russell diagram (HRD). Therefore, this relation can be used to estimate the bolometric magnitudes, luminosities, and distances of supergiants and bright giants for which only spectral information is available. If we are able to resolve massive stars in distant galaxies, this can be a very powerful tool to determine extragalactic distances.

\begin{figure}%
    \centering
    \subfloat[\centering HRD \label{fig:hr}]{{\includegraphics[width=6.3cm]{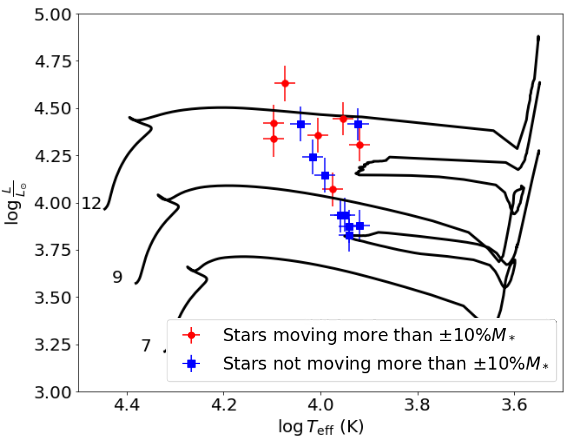} }}%
    \qquad
    \subfloat[\centering sHRD \label{fig:shr}]{{\includegraphics[width=6.3cm]{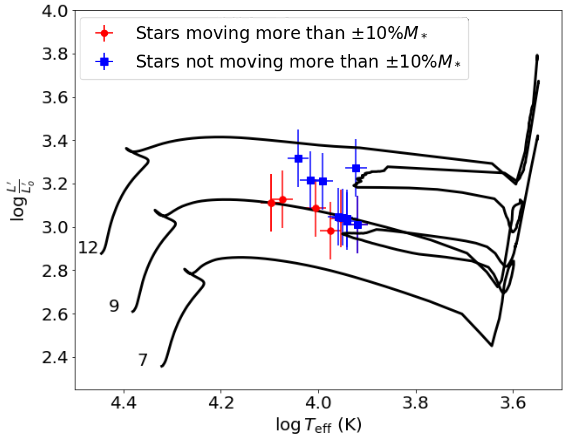} }}%
    \caption{a) HRD with evolutionary tracks for different masses (black solid lines, in $M_\odot$) accounting for rotation \citep{Evoltracks12}. The red dots represent sample stars that move by more than 10$\%$ their evolutionary mass track position with respect to the sHRD in Fig.~\ref{fig:shr}, and the blue squares are those that move less than 10$\%$. b) Same as Fig.~\ref{fig:hr}, but for the sHRD. The $L'$ is defined as the inverse of the flux-weighted gravity. %The symbol encoding refers to movements relative to the tracks with respect to Fig.~\ref{fig:hr}.
    }%
    \label{fig:example}%
\end{figure}

To study the FGLR, the stars need to be at the correct evolutionary stage. To verify that, we plotted the stars in relation to evolutionary tracks from \citet{Evoltracks12} in the regular HRD (Fig.~\ref{fig:hr}) and compared their position with respect to the same tracks in the sHRD (Fig.~\ref{fig:shr}). The sHRD \citep{Langer14} shows $L'$ which is the inverse of $g_{\mathrm{F}}$, with respect to $T_\mathrm{eff}$. Using the sHRD, we can place the stars with only their spectroscopic information and without any knowledge of their distance or brightness. The advantage of the sHRD is that the stars fall into different iso-gravity lines, enabling to discriminate stars with different radii as well as multiple sytems.

%This provides an advantage to the HRD where we need to assume a distance to obtain the luminosities. In the HRD, if two stars with same $T_{\mathrm{eff}}$ and $L$, but different masses (e.g. supergiants and post-AGB stars around $\log L/L_\odot$\,$\sim$\,4), occupy the same location, they must have the same radius, so the $\log g$ has to be different. In the sHRD, the same degeneracy does not occur since the stars fall onto different iso-gravity lines, enabling us to discriminate them. Also multiple systems, where one star dominates the observed spectrum but other objects contribute to the total luminosity, can be discriminated.

%\begin{figure}
%\centering
%\includegraphics[width=0.98\linewidth]{HRdiagram_new_axis_nohii2021_small.png}
%\caption{HRD with evolutionary tracks for different masses (black solid lines, in $M_\odot$) that account for rotation \citep{Evoltracks12}. The red dots represent sample stars that move by more than 10$\%$ their evolutionary mass track position with respect to the sHRD in Fig.~\ref{fig:shr}, and the blue squares are those that move less than 10$\%$.}
%\label{fig:hr}
%\end{figure}

%\begin{figure}
%\centering
%\includegraphics[width=0.975\linewidth]{sHRDiagram_new_axis_nohii2021_small.png}
%\caption{Same as Fig.~\ref{fig:hr}, but for the sHRD. The $L'$ is defined as the inverse of the flux-weighted gravity. The symbol encoding refers to movements relative to the tracks with respect to Fig.~\ref{fig:hr}.}
%\label{fig:shr}
%\end{figure}

%We have defined a threshold of 10$\%$ of the star mass for the change in their relative position with respect to their evolutionary tracks in both diagrams, which is reasonable since it corresponds to their mass error. 
The red dots in Figs.~\ref{fig:hr} and \ref{fig:shr} represent stars that move more than the 10\% threshold (corresponding to their mass error), which indicates that they are not well-behaved objects. These stars show significantly larger spectroscopic than evolutionary masses that are derived from comparison with evolutionary tracks.
The blue squares in Figs.~\ref{fig:hr} and \ref{fig:shr} do not move by more than the threshold, and we can assume a good correspondence between their spectral information and their true evolutionary stage and their single star status. The latter stars are certain to be in the supergiant stage and therefore the FGLR would hold. 

\begin{figure}
\centering
\includegraphics[width=.65\linewidth]{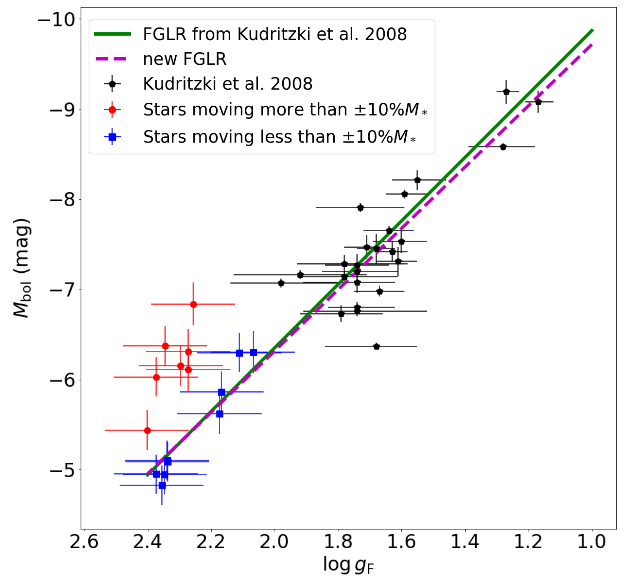}
\caption{FGLR for the stars in NGC\,300. The stars studied by \citet{Kudritzki08} are shown in black pentagons, with the corresponding FGLR regression line marked in solid green.
Blue squares and red circles mark the stars analysed in this work, with the magenta-dashed line representing the regression line derived from the black and blue symbols.}
\label{fig:myfglr}
\end{figure}

To prove our last point, we considered our 16 stars along with the
objects previously studied by \citet{Kudritzki08} to derive the FLGR.
As we can see in Fig.~\ref{fig:myfglr}, the blue squares follow the
old FGLR within their error limits. The initial FGLR gives the old parameters determined by \citet{Kudritzki08}: $a_{\mathrm{old}}$\,=\,$-3.52$ and $b_{\mathrm{old}}$\,=\,8.11. Adding the contribution of our newly found supergiants (blue squares in Fig.~\ref{fig:myfglr}), we obtained $a$\,=\,$-$3.40$\pm$0.04 and $b$\,=\,8.02$\pm$0.14. The results from this work are also in accordance with the new FGLR distance to NGC\,300 $(m-M)_{\mathrm{FGLR}}$\,=\,26.34$\pm$0.06 by \citet{2021ApJ...914...94S}.

%As previously discussed, we performed the analysis with $[Z]$\,=\,0.0\,dex and $[Z]$\,=\,$-$0.15\,dex, and although we found small systematic differences between the results, there is no observable effect on the derived flux-weighed gravities. This is due to the fact that $T_{\mathrm{eff}}$ and $\log g$ are covariant; an increase in $T_{\mathrm{eff}}$ requires an increase in $\log g$, and vice versa. Hence, $\log g_{\mathrm{F}}$ does not effectively change (in particular given the uncertainties) and nor does the FGLR since the $T_{\mathrm{eff}}$ changes are too small to introduce any effect on the derived reddening values.

\subsection{Discrepant cases}
%-------Discrepant cases--------
As seen in Fig.~\ref{fig:myfglr}, stars depicted by blue squares follow -- within the uncertainties -- the trend defined by the FGLR, while the ones represented by red dots deviate to some extent.
For the stars depicted by red circles, either the derived luminosity is too high, or the flux-weighted gravity is too large (or both). The main reason for the luminosity (bolometric magnitude) to be too high would be for the apparent magnitude to be too high. This could mean that what is seen as a single star is a combination of several unresolved sources instead. Inspecting the MUSE datacube (see Fig.~\ref{fig:photometry}), none of these objects show signs of being an extended source, which makes it unlikely that they are large star clusters. On the other hand, it cannot be ruled out that they are small stellar aggregates that are not resolved at the distance of NGC\,300. 
%We shall discuss these discrepant cases. For a star to lie above the relationship, under the consideration that it should not be there, either the derived luminosity is too high, or the flux-weighted gravity is too large (or both). The primary reason for the luminosity (bolometric magnitude) to be too high would be for the apparent magnitude to be too high. That could result from the fact that what is seen as a single star is in fact a combination of several unresolved sources. In that case, it could well happen that whilst the most luminous star dominates the observed spectrum, one or several fainter sources could be contributing to the continuum, and hence the observed apparent magnitudes. Inspection of the MUSE datacube (see Fig.~\ref{fig:photometry}) results in that none of these objects show signs of being an extended source, which makes it unlikely that they are large star clusters. On the other hand, it cannot be ruled out that they are small stellar aggregates that are not resolvable at the distance of NGC\,300. 

An alternative explanation for this deviation could be the following: as we increase the $\log g_{\mathrm{F}}$ and move to the bottom left of the FGLR, the stars decrease in mass. Population simulations predict that the FGLR will get wider for lower masses, as discussed by \citet{Meynet15}. Because of the so-called initial mass function (IMF) effect, we always expect to find a higher number of low mass stars than of massive stars \citep[e.g.][]{1955ApJ...121..161S,2001MNRAS.322..231K}, widening the FGLR because of the increased scatter.
%Nebulae around WR stars are often spectacular objects, as the intense ionizing radiation and dense, fast winds of the stellar core sweep through the remnants of the extended stellar envelope, lost either through binary stripping or winds/eruptions.
%Good examples are the nebula M1-67 around WR\,124 \citep{GroMofJon98} and NGC\,3199 around WR\,18 \citep{ToaMarGue17}.

%\begin{figure}
% \vspace*{-2.0 cm}
%\begin{center}
% \includegraphics[width=0.48\textwidth]{EM_GS96_RHD1d3d_d3l4n256_XYp45_img_004.png} 
 %\includegraphics[width=0.48\textwidth]{EM_GS96_RHD1d3d_d3l4n256_XYp45_img_016.png} 
% \vspace*{-1.0 cm}
 %\caption{Another figure}
 %  \label{fig:EM}
%\end{center}
%\end{figure}

%Here we present synthetic observations of the simulation, calculating the projected Emission Measure (EM) and X-ray surface brightness using a raytracing method, neglecting internal absorption.
%The EM is plotted in Fig.~\ref{fig:EM} as the fast wind from the WR star sweeps up the wind bubble of the previous RSG phase.

%\begin{figure}
% \vspace*{-2.0 cm}
%\begin{center}
 %\includegraphics[width=0.48\textwidth]{XR003_GS96_RHD1d3d_d3l4n256_XYp45_img_004.png} 
 %\includegraphics[width=0.48\textwidth]{XR003_GS96_RHD1d3d_d3l4n256_XYp45_img_016.png} 
% \vspace*{-1.0 cm}
 %\caption{Yet another figure}
  % \label{fig:Xray}
%\end{center}
%\end{figure}

%X-ray emission from 0.3-10\,keV is shown in Fig.~\ref{fig:Xray}, calculated using the method in \citet{GreMacHaw19}, plotted at the same times as for the EM.

%%%%%%%%%%%%%%%%%%%%%%%%%%%%%%%%%%%%%%%%%%%%%%%%%%%%%%%%%%
\section{Conclusions}
%%%%%%%%%%%%%%%%%%%%%%%%%%%%%%%%%%%%%%%%%%%%%%%%%%%%%%%%%%
We performed a quantitative spectroscopic analysis of 16 BA-type supergiants and bright giants in NGC\,300, based on VLT/MUSE IFS data. Our focus lied on determining basic atmospheric and fundamental stellar parameters. This allowed us to extend the FGLR towards less luminous stars than studied before. However, the study has faced limitations by the relatively low $S/N$\,$\lesssim$\,20 of the spectra. 
%A quantitative analysis of 16 BA-type supergiants and bright giants in NGC\,300 based on VLT/MUSE integral field spectroscopy was performed here, with a focus on determining basic atmospheric and fundamental stellar parameters. This allowed us to verify that the FGLR can be extended towards less luminous stars than studied before. However, the study has faced limitations by the rather restricted $S/N$\,$\lesssim$\,20 of the spectra. 
For future work, the $S/N$ and spatial resolution should be improved by taking advantage of the adaptive optics mode of MUSE and longer exposure times. This would not only reduce the uncertainties for similar studies as this work, but would also help to determine metallicities and likely elemental abundances for selected individual chemical elements. Therefore, demonstrating the full potential of MUSE for extragalactic stellar astrophysics. In addition, BlueMUSE is the new proposed medium-resolution IFS instrument at the VLT. Optimized for the optical blue, BlueMUSE will be the perfect instrument to study hot massive stars as most of their spectral features are located in its spectral range - a Highlight Science Case outlined in the BlueMUSE White  Paper\citep{2019arXiv190601657R}. 

%longer exposures and use of the currently available AO mode of MUSE
%should be aimed at boosting the $S/N$ and improving the spatial resolution. This would not only facilitate a similar study as the one performed here to be achieved at much reduced uncertainties,
%but this would also help to determine metallicities and likely elemental abundances for selected individual chemical elements. The full potential of MUSE for extragalactic stellar astrophysics could thus be demonstrated. %Studies with increased scientific value could be made once
%the BlueMUSE medium-resolution panoramic integral field spectrograph becomes operational, which has been proposed as a new instrument for the VLT \citep{2019arXiv190601657R}. BlueMUSE is optimised for the optical blue, where most of the diagnostic spectral features of hot supergiants are located.

%%%%%%%%%%%%%%%%%%%%%%%%%%%%%%%%%%%%%%%%%%%%%%%%%%%%%%%%%%
%\section*{Acknowledgements}
%%%%%%%%%%%%%%%%%%%%%%%%%%%%%%%%%%%%%%%%%%%%%%%%%%%%%%%%%%
%Thanks to everyone for their help.

%%%%%%%%%%%%%%%%%%%%%%%%%%%%%%%%%%%%%%%%%%%%%%%%%%%%%%%%%%
\bibliographystyle{apj}
\bibliography{./sample2}
%%%%%%%%%%%%%%%%%%%%%%%%%%%%%%%%%%%%%%%%%%%%%%%%%%%%%%%%%%

%%%%%%%%%%%%%%%%%%%%%%%%%%%%%%%%%%%%%%%%%%%%%%%%%%%%%%%%%%
\begin{discussion}

\discuss{Question}{
How do you know that the stars are in the right evolutionary stage?
}

\discuss{Gonz\'alez-Tor\`a}{We plot all the targets both in the HRD and the sHRD. The blue dots in Figures~\ref{fig:hr}, \ref{fig:shr} are the stars in the supergiant phase, since their photometric and spectroscopic information coincide \textit{(see more in Section~\ref{sec:fglr}, 3rd and 4th paragraphs).}
 % It shouldn't -- it depends how symmetric the Riemann solver is.  If it is perfectly symmetric then all of the octants should be identical.  At the moment I am looking into adding a stochastic model for clumpiness to the RSG wind, because we know that the winds of cool stars are very clumped. We expect these non-linear clumps in the wind will then seed all of the structure that we see in the WR nebulae.
}

%\discuss{Mohamed}{
%  Could you say something about the computational cost of the TORUS post-processing of PION simulations.
%}

%\discuss{Mackey}{
%  At the beginning the TORUS calculations were taking longer than the hydro simulations, but Tom and Sam did a lot of work on this.
%  In the end with MPI and multi-threading, the TORUS calculations took about 10\,000 core hours for a 3D simulation that required about 50\,000 core hours.
%}
\end{discussion}
%%%%%%%%%%%%%%%%%%%%%%%%%%%%%%%%%%%%%%%%%%%%%%%%%%%%%%%%%%

\end{document}